\documentclass[12pt]{article}
\usepackage{graphicx}

\begin{document}

\begin{center}
\Large {\bf  Physical effects of massless cosmic strings}
\end{center}

\bigskip
\bigskip

\begin{center}
D.V. Fursaev
\end{center}

\bigskip
\bigskip

\begin{center}
{\it Dubna State University \\
     Universitetskaya str. 19\\
     141 980, Dubna, Moscow Region, Russia\\

  and\\

  the Bogoliubov Laboratory of Theoretical Physics\\
  Joint Institute for Nuclear Research\\
  Dubna, Russia\\}
 \medskip
\end{center}

\bigskip
\bigskip

\begin{abstract}
We study massless cosmic strings which are one-dimensional objects moving with the speed of light. 
Perturbations of velocities of test bodies and anisotropy of cosmic microwave background generated by massless cosmic strings are analysed.  These phenomena are analogous to the string wake effect and the Kaiser-Stebbins effect for standard (massive) cosmic strings. There are two regimes depending on the energy $E$ of a massless string per unit length. At low energies, $EG/c^4\ll 1$, massless and massive strings act similarly. At high energies, $EG/c^4 \sim 1$, massless string effects are quite different.
Our work provides a method to describe different physical phenomena on spacetimes of massless strings which take into account the presence of a parabolic holonomy.
\end{abstract}

\newpage

\section{Introduction}\label{intr}

Cosmic strings (CS) are macroscopic one-dimensional objects which are believed to appear during phase 
transitions in the early 
Universe \cite{Kibble:1976sj}.  Physical effects in the gravitational 
field of CS were a subject of intensive studies \cite{Vilenkin:2000jqa} since CS were considered as a possible source of primordial density fluctuations. 
CS are specified by a tension $\mu$ which determines the mass per unit length in the rest frame.  

A massive CS does not produce a local 
gravitational field. Its effect is global, in a sense that a length of a unit circle
around a string at rest is less than $2\pi$ with the angle deficit being $8\pi G \mu/c^2$.
This means that there is a non-trivial holonomy for a closed contour around 
the string in a flat space-time. The trajectories of test bodies or light rays passing from one side of the string are transformed by the holonomy with respect to
trajectories passing the string from the other side. A moving  CS generates perturbations of the velocities of bodies resulting in the so called the string wake effect. The moving string also perturbs frequencies of quanta, and yields an  
anisotropy of cosmic microwave background known as the Kaiser-Stebbins effect \cite{Stebbins:1987va}. Together with lensing effects these physical phenomena are used in methods of CS observational search \cite{Sazhina:2008xs},\cite{Brandenberger:2013tr}.

In this paper we distinguish massive and massless cosmic strings. Massive CS are just standard cosmic strings described above.  They are specified by a nonvanishing mass
per unit length in a rest frame. Therefore, their velocity can never reach the speed of light. The rest of the paper is devoted to one-dimensional string-like objects which move with the speed of light.   Their background geometry can be obtained from 
the metric of a moving massive string \cite{Barrabes:2002hn} as a result of the Aichelburg-Sexl boost, when the velocity reaches the speed of light, mass tends to zero, while energy remains finite. We call such objects massless cosmic strings (MCS). Massless strings are specified in a given frame of reference by energy $E$ per unit length. 

String-like microscopical objects which move with the speed of light were first considered in  1970's when quantized string states were used to describe some resonances
mediating hadronic interactions. They were called null strings \cite{Schild:1976vq}. Equations of motion of null strings suggest that their worldsheet is a geodesic null 
hypersurface, i.e. each point of the string moves along a null geodesic.  

In this paper, we study macroscopic MCS and address the following question:
if massless cosmic strings existed in Nature what would be their specific features 
from the point of view of observations? Physical effects caused by MCS have not been studied so far. We work in a locally flat 
spacetime and show that MCS allow analogues of the string wake 
and the Kaiser-Stebbins effects.

It should be noted that dealing with the geometry of a massless CS is less trivial 
than in the massive case. The Aichelburg-Sexl boost yields a metric whose
components behave as distributions. It was shown in \cite{vandeMeent:2012gb} that such a metric does not result in gravitational 
shockwaves since the background geometry is smooth everywhere except singularities (analogous conical singularities) located on the string worldsheet. The singularities are determined by a holonomy which belongs to the subgroup of so called parabolic Lorentz transformations, 
a parabolic holonomy. Our results reveal an operational meaning of MCS spacetime. The computations can be done in terms
of the holonomy transformations applied at the string event horizon. 
 
The paper is organized as follows. In Section 2 we describe the metric of a massless CS 
moving in a flat spacetime and its parabolic holonomy. Section 3 presents a method
to study geodesics in the gravitational field of the string. We introduce a regularization
of delta-function singularities in the metric by considering a consequence of strings
with a smooth profile. After regularization is removed, geodesics passing event horizon of the string exhibit a discontinuity with respect to the coordinate chart.
They are transformed by the square root of the holonomy. Mutual rotation of trajectories passing from different sides of the string is determined by the complete holonomy, as expected.
Physical effects in the gravitational field of the string are considered in Section 4
(wake effects) and in Section 5 (Kaiser-Stebbins effect).
Computations show that MCS at high energies can be definitely distinguished in observations from massive CS. A precise formula of frequency shifts of quanta and anisotropy
of cosmic microwave background is determined in Section 5. An interesting property of CMB anisotropy caused by MCS is that it is located on a CMP map inside a single spot.
The boundary of the spot is an image of the string formed by photons emitted toward
the observer from the string. Section 6 contains a summary of our method how to deal with MCS spacetimes. The key idea is that the string event horizon can be considered as a kind of a Cauchy hypersurface for phenomena above the horizon.
The Cauchy data from one side of the string worldsheet on the horizon should be transformed by the holonomy w.r.t. the Cauchy data on the other side of the horizon.

\section{Metric around a massless cosmic string}\label{mls}
\setcounter{equation}0

We start with a massive CS with a tension $\mu$ which is at rest in a locally Minkowsky spacetime.
If the string is parallel along the $z$-axis its metric can be written as:
\begin{equation}\label{2.1}
ds^2=-dt^2+dz^2+dx^2+dy^2+(\alpha^2-1){(xdy-ydx)^2 \over x^2+y^2}~~,
\end{equation}
where $\alpha=1-4\mu G$ (for the velocity of light we put $c=1$). In (\ref{2.1})
we use a Cartesian like coordinate, instead of common polar coordinates. Coordinates
$x$ and $y$ are defined on the entire plane except the point $x=y=0$, where the string
is located. 

The Ricci tensor for (\ref{2.1}) is $R_{\mu\nu}=2\pi(1-\alpha)\delta(x)\delta(y)
n^i_\mu n^i_\nu$, where $n^i$ is a pair of unit vectors orthogonal to the string worldsheet
\cite{Fursaev:1995ef}. Let $l$ be a unit vector along the string, $u$ be a vector of its four velocity. By using the above form the the Ricci tensor one can easily see that the Einstein equations are fulfilled for the following stress-energy tensor of the string 
\begin{equation}\label{2.2}
T_{\mu\nu}=\rho (u_\mu u_\nu -l_\mu l_\nu)~~,
\end{equation}
with $\rho=\mu \delta(x)\delta(y)$. Thus, the pressure along the string equals minus its energy.

Metric for a massless cosmic string can be obtained from (\ref{2.1}) by using the Aichelburg-Sexl boost (the Penrose limit) in the following way. One considers (\ref{2.1}) in a frame where the string is moving with a velocity $v$ along, say,
the $x$ axis. The relation between coordinates is $x=\gamma(x'-vt')$, where $t'$ and $x'$
are coordinates in the frame where the string is moving, $\gamma=1/\sqrt{1-v^2}$. In the
moving frame one goes to the limit $v\to 1$, $\mu\to 0$, such that the energy $E$ per unit length remains finite,
\begin{equation}\label{2.3}
E=\lim_{v\to 1} (\gamma \mu)~~.
\end{equation}
In the light cone coordinates, $u=t'-x'$, $v=t'+x'$, (\ref{2.1}) takes the form \cite{Barrabes:2002hn}:
\begin{equation}\label{2.4}
ds^2=-du dv+dz^2+d y^2- 8\pi G E |y| \delta(u) du^2~~.
\end{equation}
To get (\ref{2.4}) one should use the fact that function $\gamma/(x^2+y^2)$ converts to
the distribution $\pi \delta(u)/|y|$. 

Element (\ref{2.4}) describes a metric on a spacetime of MCS. The string worldsheet  
is a geodesic null hypersurface given by equations $u=y=0$. The problem with (\ref{2.4}) is that it has singularity on a larger null
hypersurface $u=0$. This hypersurface is an event horizon of the string. All events
which happen above the horizon, $u>0$, cannot affect the string. Since the string itself 
moves along the horizon, all events below the horizon, $u<0$, are casually independent of
string. The singularity at $u=0$, $y\neq 0$ looks as a shock wave. We show 
in Section 3 that it is not the case: there are no singularities
in the components of the Riemann tensor on the string horizon, and the local geometry 
outside  the string is flat.

The boost applied to (\ref{2.2}) yields the following stress-energy tensor of MCS:
\begin{equation}\label{2.5}
T_{\mu\nu}=\tilde{\rho} u_\mu u_\nu~~,
\end{equation}
where $\tilde{\rho}=E\delta(y)\delta(u)$ is the energy density of the string and $u^\mu$ 
is its 4-velocity, $u^2=0$. The 4-velocity of MCS follows 
from the 4-velocity of the massive string multiplied by factor $\gamma$. Note that $x/\gamma$
transforms to the null coordinate $u$, $\gamma \delta(x)$ transforms to $\delta(u)$, and
$\gamma^2\rho$ transforms to $\tilde{\rho}$. Equation (\ref{2.5}) show that parameter $E$ in (\ref{2.4}) does determine the energy of the string.

Since the geometry around the string is locally flat except singularities located on the  string worldsheet, there must exist a non-trivial holonomy when going around the string \cite{vandeMeent:2012gb}. Let us start with the holonomy around a massive string at rest.
It belongs to a rotation subgroup of the Lorentz group, i.e. rotations in the $x,y$ plane 
around the string by angle $\alpha=1-4\mu G$. It is convenient to write the
corresponding matrices 
in terms of veirbains:
\begin{equation}\label{2.6}
O(\alpha)=I+(\cos\alpha -1)(p_x \otimes p_x + p_y \otimes p_y)\eta
+\sin\alpha ~(p_x \otimes p_y - p_y \otimes p_x)\eta~~.
\end{equation}
Here $\eta$ is the flat metric tensor, $p_x$, $p_y$ are two unit mutually orthogonal vectors directed along $x$ and
$y$ coordinate axes. In component notations, 
$(a \otimes b~\eta)^\mu_\nu=a^\mu b_\nu$. The same holonomy in the moving frame is
given by matrix $O'(\alpha)$ obtained from $O(\alpha)$ by 
substituting $p_x$ to $p'_x=\gamma(p_x+v p_0)$, where $p_0$ is future directed time-like vector, $p_0^2=-1$.
The Aichelburg-Sexl boost applied to $O'(\alpha)$ yields the holonomy around
the MCS:
\begin{equation}\label{2.7}
M(\lambda)=\lim_{v\to 1} O'(\alpha)=I-\left({\lambda^2 \over 2}u \otimes u +\lambda p_y \otimes u -\lambda u \otimes p_y\right)\eta~~,
\end{equation}
$$
\lambda\equiv 8\pi GE~~.
$$ 
Here $u=p_0+p_x$ is the 4-velocity vector of the string. 
It is easy to check that $M(\lambda_1)M(\lambda_1)=M(\lambda_1+\lambda_2)$. Thus, $M(\lambda)$ make a one-parameter (hyperbolic) subgroup of the Lorentz transformations. It is easy to check that
\begin{equation}\label{2.7a}
M(\lambda)u=u~~,~~M(\lambda)l=l~~,~~M(\lambda)v=v+\lambda^2 u+2\lambda p_y~~,
~~M(\lambda)p_y=p_y+\lambda u~~,
\end{equation}
where $l$ is directed along the string, $v=p_0-p_y$. 
Since $M(\lambda)$ leave $u$ invariant they are also called null rotations. 

Matrix (\ref{2.7}) determines transformation of a vector $a$ under a parallel transport around  the string, $a\to a'=M(\lambda)a$. Transformations of components of this vector in 
the veirbain basis $p_0$, $p_x$, $p_y$, $l=p_z$ are:
\begin{equation}\label{2.8a}
(a')^0=\left(1+{\lambda^2 \over 2}\right)a^0-{\lambda^2 \over 2}a^x+\lambda a^y~~,
\end{equation}
\begin{equation}\label{2.8b}
(a')^x=\left(1-{\lambda^2 \over 2}\right)a^x+{\lambda^2 \over 2}a^0+\lambda a^y~~,
\end{equation}
\begin{equation}\label{2.8c}
(a')^y=a^y+\lambda(a^0-a^x)~~,~~(a')^z=a^z~~.
\end{equation}
Here we used the relation $a'=a^b p'_b=(a')^bp_b$. For the light cone components,
$a^u=a^0-a^x$, $a^v=a^0+a^x$, one has 
\begin{equation}\label{2.8d}
(a')^u=a^u~~,~~(a')^v=a^v+\lambda^2 a^u+2\lambda a^y.
\end{equation}

As we see in Section 3, transformations (\ref{2.8a})-(\ref{2.8d}) determine physical
effects in the gravitational field of the massless string. Since effects are time dependent,
some other information in addition to the holonomy is needed.

We complete this Section with a remark that energy $E$ of the string, like the energy
of a photon, depends on a frame of reference where it is measured. Therefore, different 
$E$ in (\ref{2.4}) are not different sorts of massless strings (as opposed to massive strings whose tensions $\mu$ at rest allow one to distinguish their origin).
Transformation of $E$ under boosts can be extracted from (\ref{2.7}).
If another frame moves along the $x$ axis with the coordinate velocity $v_1$ with respect to coordinates (\ref{2.4}) the holonomy in the new frame is given by matrix (\ref{2.7}), where vector  $u$ should be replaced with $u'=e^\beta u$, $\tanh \beta=v_1$. This is the same as if
$u$ is left unchanged and $E$ is replaced to $E'=e^\beta E$. The last relation is the standard Doppler effect.

\section{Constructing trajectories in the string spacetime}\label{reg}
\setcounter{equation}0
 
Singularities in (\ref{2.4}) outside the string worldsheet are coordinate singularities.
They appear since there is non-trivial holonomy (\ref{2.7}) which is ignored by a coordinate chart 
used in (\ref{2.4}). We expect that in these coordinates physical trajectories experience discontinuities when passing the MCS horizon.

Let us demonstrate that there are no curvature singularities on the string horizon outside the string. We replace (\ref{2.4}) 
with the following 'regularized' metric:
\begin{equation}\label{3.1}
ds^2=-du dv+dz^2+d y^2- \lambda |y| \chi(u) du^2~~,
\end{equation}
where $\chi(u)$ is a smooth function with a finite support inside the interval
$-\epsilon <u\leq 0$, $\epsilon>0$.

Computations show that the Riemann tensor of (\ref{3.1}) has the only non-vanishing component
$R_{uyuy}=\lambda \chi(u) \delta(y)$. When the regularization 
is removed, $\chi(u)\to \delta(u)$,  singularities appear on the string worldsheet.
One can also examine the Einstein tensor to see that (\ref{3.1}) is a solution to the Einstein 
equations with the stress-energy tensor having the single non-vanishing component $T_{uu}=E\chi(u) \delta(y)$.
This can be interpreted as a source of a 'pulse of massless strings' moving along the $x$ axis. 
The energy profile in the pulse is $E\chi(u)$.

Geodesics in metric (\ref{2.4}) are locally straight lines except 
the horizon $u=0$. In (\ref{3.1}) the horizon discontinuities are regularized
and transition of trajectories across the horizon can be investigated by standard methods.

Geodesic equations for the components of 4-velocity $u^\mu=dx^\mu/d\sigma$ look as follows
\begin{equation}\label{3.2d}
{d u^y \over d\sigma}=\frac 12 g_{,y}(u^u)^2~~,
\end{equation}
\begin{equation}\label{3.2e}
{d u^v \over d\sigma}={dg \over d\sigma}u^u+g_{,y}u^u u^y~~.
\end{equation}
Here $g=g(u,y)=-\lambda |y|\chi(u)$, and $\sigma$ is an affine parameter.
Components $u^u$ and $u^z$ remain constant along the trajectory. 

It is convenient to use the relation $d\sigma = du/\bar{u}^u$ and consider  components as functions of the $u$ coordinate, $u^\mu=u^\mu(u)$. Let $\bar{u}^\mu$ be components before the trajectory crosses the horizon, that is $\bar{u}^\mu=u^\mu(-\epsilon)$.
Integration of (\ref{3.2d}),(\ref{3.2e}) yields:
\begin{equation}\label{3.2a}
u^u=\bar{u}^u~~,~~u^z=\bar{u}^z~~,
\end{equation}
\begin{equation}\label{3.2b}
u^y(u)=\bar{u}^y+ {\bar{u}^u \over 2} \int^u_{-\epsilon}g_{,y}(u',y(u'))du'~~~,
\end{equation}
$$
u^v(u)=\bar{u}^v+g(u,y(u))\bar{u}^u+\bar{u}^y\int^u_{-\epsilon}g_{,y}(u',y(u'))du'+
$$
\begin{equation}\label{3.2c}
{\bar{u}^u  \over 2} \int^u_{-\epsilon}g_{,y}(u',y(u'))du'\int^{u'}_{-\epsilon}g_{,y}(u'',y(u''))du''~~~.
\end{equation}
After  regularization is removed
(\ref{3.2a})-(\ref{3.2c}) acquire a simple form above the horizon ($u>0$).
By comparing these relations with (\ref{2.8c}),(\ref{2.8d}) 
one comes to the following relation between 4-velocities
before  and after crossing the horizon $u>0$:
\begin{equation}\label{3.3}
u^\mu=M^\mu_{~\nu}\left(\mp \lambda /2 \right)\bar{u}^\nu~~.
\end{equation}
Here signs "-" and "+" correspond to trajectories with coordinates $y>0$ and $y<0$, 
respectively. Thus, geodesics change its orientation with respect to the
coordinate chart after crossing the horizon. This change is given by a square root 
of the holonomy around the string (since it is determined by $\lambda/2$). Transformations of trajectories which bypass the string from different sides are just 
opposite.

Note that $u^v$ also has a contribution $-\lambda |y|\chi(u)$, see (\ref{3.2b}),
which signals that the $v$-coordinate of the trajectory makes a jump on the horizon
$\Delta v=-\lambda |y|$. By taking into account (\ref{2.8a})-(\ref{2.8d}) one can put it in the same form as (\ref{3.3}): the position of the trajectory leaving
the horizon has coordinates  $x^\mu=M^\mu_{~\nu}\left(\mp\lambda/2\right)\bar{x}^\nu$, where $\bar{x}^\mu$ are coordinates with which 
it enters the horizon.

For physical applications it is important to know transformation of the trajectories
with respect to each other rather than with respect to the coordinate chart (\ref{2.4}). It is clear
from (\ref{3.3}) that trajectories being on one side of the string worldsheet do not 
experience mutual transformation. In particularly, this property shows that string produces no shockwaves.

One can introduce a coordinate chart, different from (\ref{2.4}), which 
behaves smoothly across the horizon, say for $y>0$. One can call it a 'right' chart
or $R$-chart, for brevity.
In $R$-chart  geodesics crossing the horizon at $y>0$ ('right' geodesics) are straight lines without discontinuities. The 'left' geodesics which cross the horizon at $y<0$ change their position and the direction. Thus, $R$-chart is singular on the 
left part of the horizon, $u=0, y<0$. Suppose that $\bar{x}^\mu$ and $\bar{u}^\mu$ are the data of the 
'left' trajectory  with which it approaches the horizon from $u<0$. Then  initial data on the R-chart for the 'left' trajectory leaving the horizon are
\begin{equation}\label{3.4}
x^\mu=M^\mu_{~\nu}\left(\lambda\right)\bar{x}^\nu\mid_{u=0,y<0}~~,
~~u^\mu=M^\mu_{~\nu}\left(\lambda\right)\bar{u}^\nu\mid_{u=0,y<0}~~.
\end{equation}  
This relation follows from (\ref{3.3}) since both the 'left' geodesic and the R-chart are transformed with respect to the coordinate chart in (\ref{2.4}). The mutual transformation is given by the complete holonomy, as was expected.

The subsequent description of physical effects in next Sections is based on the conclusion that 'right' observers (who are at rest with respect to the $R$-chart) meet 'left' geodesics coming from the horizon with  data transformed according to (\ref{3.4}). 
Obviously one could introduce 'left' coordinates ($L$-charts), where 'left' geodesics are 
always straight lines. In $L$-charts 'right' geodesics experience transformation on the horizon of the form (\ref{3.4}), where $\lambda$ should be replaced with $-\lambda$.

As a simple exercise, consider 'left' and 'right' observers who are initially at rest with respect to each other, and a massless string moving exactly in between. Let $(\bar{x}_R=0,\bar{y}_R=a,\bar{z}_R=b)$, $(\bar{x}_L=0,\bar{y}_L=a,\bar{z}_L=b)$ be their coordinates ($a>0$) before crossing the horizon. 
The 'right' observer crosses the horizon at the moment $t_R=0$.
The 'left' observer (from the point of view of the 'right' one) appears from the horizon at the moment $t_L=t_H=-\lambda a$ with coordinates $x_L=-\lambda a,y_L=-a,z_L=b$, see (\ref{2.8a})-(\ref{2.8d}). Since its 4-velocity is $u_L^0=1+\lambda^2/2$, $u_L^x=\lambda^2/2$, $u_L^y=\lambda$ (the original velocity being $\bar{u}_L^\mu=\delta^\mu_0$) the 'left' observer moves to the right with the coordinate velocity having components:
\begin{equation}\label{3.5}
v^x_L={\lambda^2 \over 2+\lambda^2}~~,~~v^y_L={2\lambda \over 2+\lambda^2}~~.
\end{equation}
The trajectory of the 'left' observer in the coordinates, where the 'right' observer is at 
rest, is $x_L(t)=-\lambda a+v^x(t-t_H)$, $y_L(t)=- a+v^y(t-t_H)$, where $t$ is a time coordinate which coincides with the proper time of the 'right' observer. 
Therefore, the observers meet at the moment 
$$
t=t_H+{2a \over v^y}={L \over 2\lambda}(\lambda^2-\lambda+2)~~,
$$
where $L=2a$ is the distance between the observers.
For low-energy strings $EG\ll 1$ this time is $t\simeq L/(EG)$. If $\lambda \sim 0.1$, 
observers located in the same galaxy  at a distance, say, $10^4$ light years may meet
after $10^5$ years.

If instead of the above two observers one had two particles of equal masses $m$, they would 
collide with the total energy in the center of mass $E=(2+\lambda^2/4)m$. High energy strings $EG\simeq 1$
can 'accelerate' particles up to energies about $160~ m$.
 
\section{Wake effects}\label{wake}
\setcounter{equation}0

\begin{figure}[h]
\begin{center}
\includegraphics[height=8cm,width=15cm]{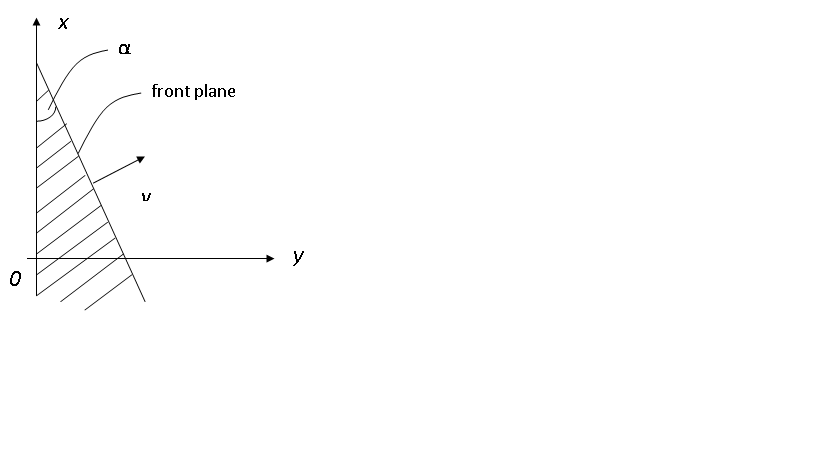}
\caption{\small{Dashed region is a wedge-shaped overdensity as measured by observers
at $y>0$. Overdensities are caused by perturbations of velocities of the matter at $y<0$ generated by a massless string moving along the $x$ axis at $y=0$. The front of the wedge moves together with the string. The tilt of the front to the string trajectory is $\alpha$.}}
\label{F4}
\end{center}
\end{figure}
 
Although MCS do not produce shock waves they generate perturbations of
mutual velocities of bodies. As a result, domains of matter which were initially at rest  may move to each other after crossing the horizon and collide or overlap. This is known as a wake effect.

Let $\bar{v}^i=\bar{u}^i/\bar{u}^0$ be coordinate velocity of a left trajectory. 
According to (\ref{2.8a})-(\ref{2.8d}), its velocity as measured by a 'right' observer 
becomes 
\begin{equation}\label{4.1a}
v^x={\left(1-{\lambda^2 \over 2}\right)\bar{v}^x+{\lambda^2 \over 2}+\lambda \bar{v}^y
\over \left(1+{\lambda^2 \over 2}\right)-{\lambda^2 \over 2}\bar{v}^x+\lambda \bar{v}^y}~~,
\end{equation}
\begin{equation}\label{4.1b}
v^y={\bar{v}^y+\lambda(1-\bar{v}^x)\over \left(1+{\lambda^2 \over 2}\right)-{\lambda^2 \over 2}\bar{v}^x+\lambda \bar{v}^y}~~,
\end{equation}
\begin{equation}\label{4.1c}
v^z={\bar{v}^z \over \left(1+{\lambda^2 \over 2}\right)-{\lambda^2 \over 2}\bar{v}^x+\lambda \bar{v}^y}~~.
\end{equation}
For non-relativistic trajectories $\bar{v}^i\ll 1$ and low-energy strings $\lambda \ll 1$ perturbations of the velocity $\delta v^i=v^i-\bar{v}^i$ in the leading order look as:
\begin{equation}\label{4.2}
\delta v^x=\lambda v^y~~,~~\delta v^y=\lambda~~,~~\delta v^z=0~~.
\end{equation}
Therefore, the main effect is that velocity changes in the direction orthogonal 
to the string world-sheet by the quantity $\delta v=\delta v^y\simeq 8\pi GE$. 

It is very similar to the perturbation of the velocity by a moving massive string
$\delta v=4\pi G v_s\gamma(v_s) \mu$, where $v_s$ is the velocity of the string
\cite{Brandenberger:2013tr}. In the ultra relativistic limit, 
$v_s\to 1$, $\gamma(v_s)\mu\to E$, this yields $\delta v=4\pi GE$.

Behind the moving strings there appear regions with over densities \cite{Brandenberger:2013tr}. Let us see how 
such a regions form behind a massless string. Suppose that the matter is distributed 
over the spacetime with a constant density $\rho$, and it is at rest below the string horizon. Above the string horizon 'left' matter (at $y<0$) starts to move to the right, and doubles the density in a region at $y>0$. On the $R$-chart the region with over density has a wedge-like shape, see Figure \ref{F4}.
At the moment of time $t$ (as measured by a 'right' observer) it consists of a half plane
($y=0$, $x\leq t$) and a front plane. The front plane is formed by density perturbations
which move to right just from the very vicinity of the string worldsheet ($x=t,y=0$).

In $(x,y)$ plane trajectory of a perturbation which starts to move from $y=0$ at a moment $t=\eta$ is
\begin{equation}\label{4.3}
x(t)=\eta+v_x(t-\eta)~~,~~y(t)=v_y(t-\eta)~~,
\end{equation}
$v^i$ are defined in (\ref{3.5}).
The equation of the front at a moment $t$ can be obtained from (\ref{4.3}) in the
following form
\begin{equation}\label{4.4}
x(y;t)={v_x-1 \over v_y}y+t~~,~~y>0~~.
\end{equation}
Equation (\ref{4.4}) is a half plane.

The front is tilted to the direction of the string movement ($x$ axis) at angle $\alpha$
such that 
\begin{equation}\label{4.5}
\cot \alpha={v_y \over 1-v_x}=\lambda.
\end{equation}

The front reaches a right observer with coordinates $x,y,z$
at the moment $t=x-(v_x-1)y/v_y$. The front moves with the velocity $v=\sin \alpha=
1/\sqrt{1+\lambda^2}$.

\section{Optical effects and CMB anysotropy}\label{hc}

Massless strings perturb 4-momenta of rays crossing the string horizon.
This results in Doppler-like shifts in the energy of rays. Now 
we describe how these effects are seen by observers.

\begin{figure}[h]
\begin{center}
\includegraphics[height=8cm,width=11cm]{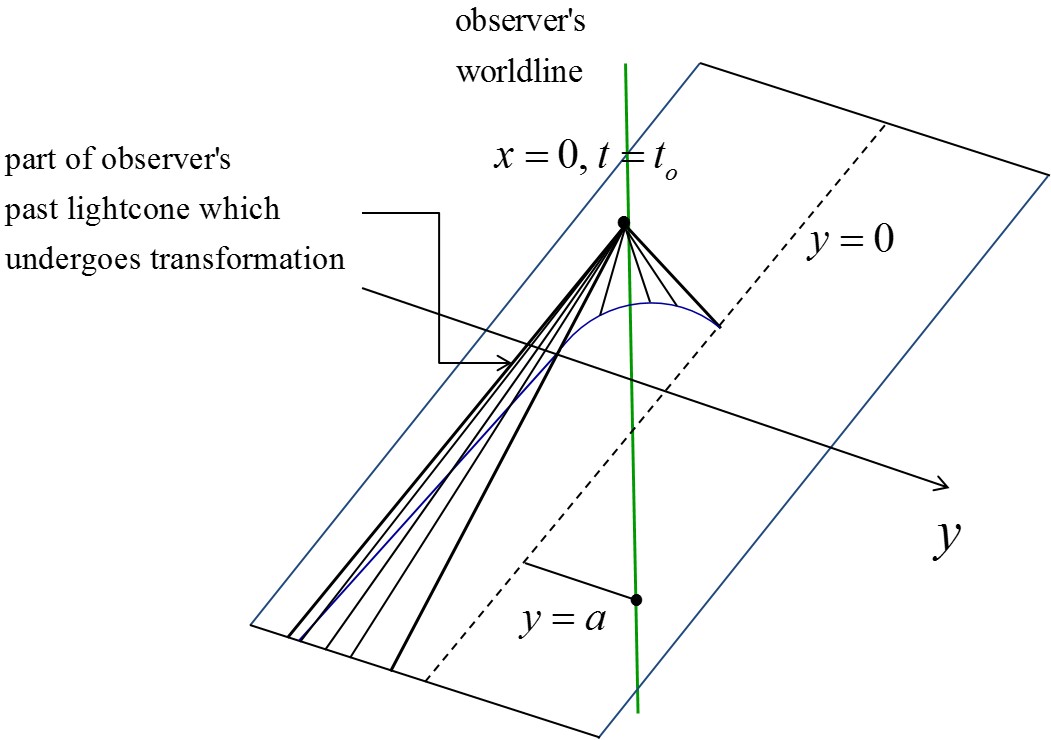}
\caption{\small{shows trajectories of 'left' rays which are registered
by a 'right' observer. The trajectories make a part of the past light cone which
starts from the horizon at $y<0$.}}
\label{F1}
\end{center}
\end{figure}

Consider a 'right' observer which is at rest at some fixed distance to the string
trajectory. We are interested in 'left' rays he can detect. At a given moment, $t=t_o$, trajectories
of these rays make a part of observer's past light cone, see Figure \ref{F1}.
The boundary between 'left' and 'right' parts are rays which start at string trajectory $y=0$. These boundary rays make an image of the sting seen by the observer.

The equation for the past light cone for an observer with coordinates ${\bf x}_o$ at a moment $t_o$ is 
\begin{equation}\label{5.1}
{\bf x}(t)={\bf x}_o+(t-t_o){\bf n}~~,~~{\bf n}^2=1~~,~~t<t_o~~.
\end{equation}
At the moment of emission from the string horizon, $t=t_e$, the 'left'  rays are specified by conditions $x(t_e)=t_e$, $y(t_e)<0$.  For rays which make an image of the string $y(t_e)=0$.
It is convenient to fix direction of $\bf n$ by spherical angles $\theta$ and $\varphi$ 
as $n_x=\sin\theta \cos\varphi$, $n_y=\sin\theta \sin\varphi$,
$n_z=\cos\theta$,  $0<\theta<\pi$, and $0<\varphi \leq 2\pi$.

Suppose the position of the observer in the $(x,y)$ plane is $x_o=0$, $y_o=a$, where $a>0$. 
Position along 
the $z$-axis is fixed, the actual value of the $z$-coordinate does not matter.   
To get the direction of the unit vector ${\bf n}$ which corresponds to the string image one should put $y(t_e)=0$ in (\ref{5.1}). This yields
\begin{equation}\label{5.2}
\sin^2\theta={t_e^2+a^2 \over (t_o-t_e)^2}~~,~~\cot \varphi=-{t_e \over a}~~,
\end{equation}
The time of emission $t_e$ can be excluded from (\ref{5.2}), and one gets the following 
formula:
\begin{equation}\label{5.3}
\sin\theta\sin(\varphi + \bar{\theta})=\sin \bar{\theta}~~,
\end{equation}
\begin{equation}\label{5.4}
\sin\bar{\theta}={a \over \sqrt{a^2+t_o^2}}~~.
\end{equation}
We assume here that $0<\bar{\theta}<\pi/2$. 

\begin{figure}[h]
\begin{center}
\includegraphics[height=8cm,width=9.5cm]{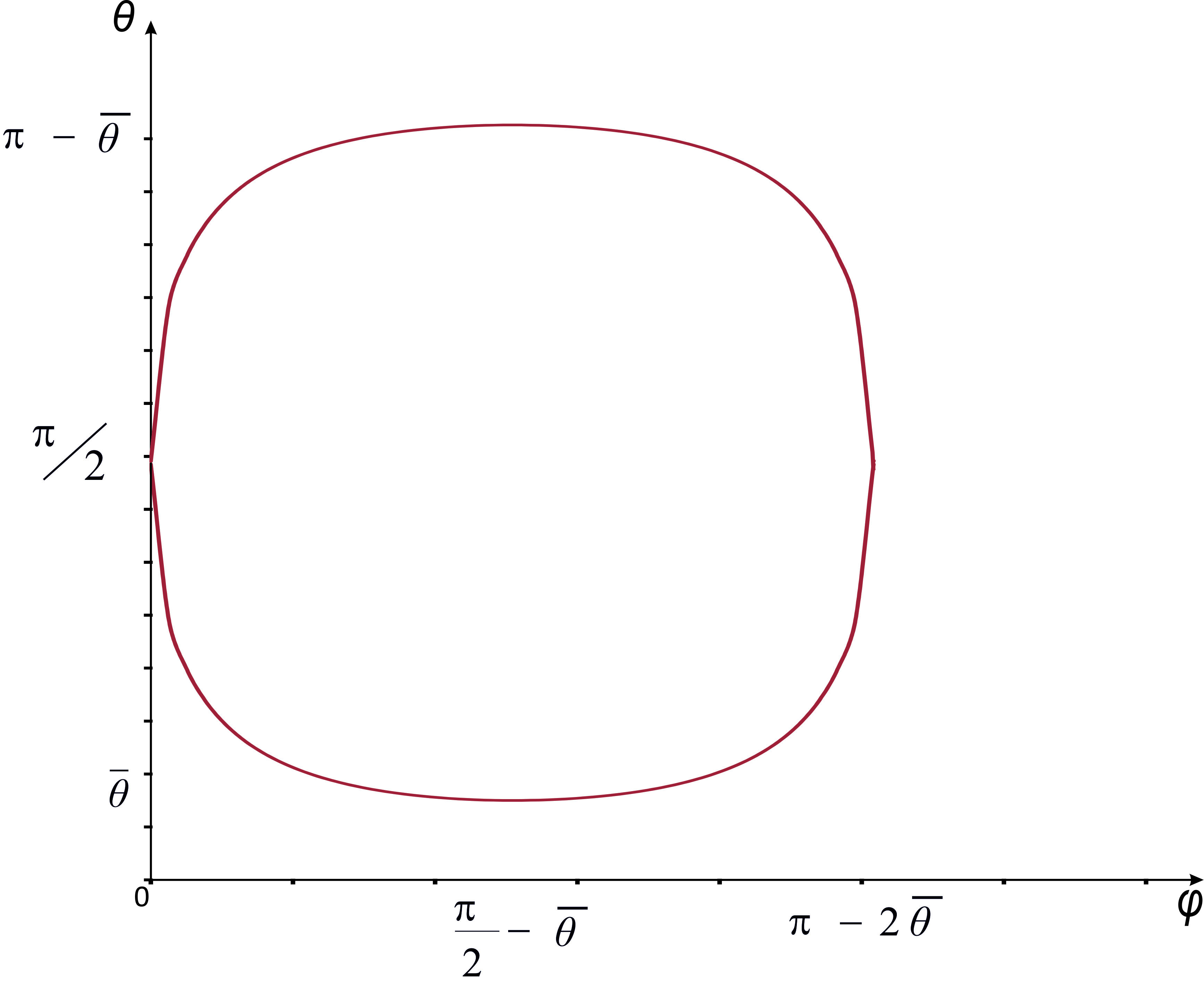}
\caption{\small{shows the image of the string described in spherical coordinates by Eq. (\ref{5.3}).}}
\label{F2}
\end{center}
\end{figure}

The observer sees the string in the spherical coordinates $\theta,\varphi$ as an ellipse-like curve, see Figure \ref{F2}. Points on the curve correspond to light rays emitted 
at different moments by different parts of the string. For instance, the point $(\theta=\pi/2,\varphi=0)$ corresponds to rays
emitted in the far past, $t_e\to -\infty$, while the opposite point $(\theta=\pi/2,\varphi=\pi - 2\bar{\theta})$ is a ray emitted at $t_e=t_e^{max}=(t_o-a^2/t_o)/2$ in the direction orthogonal to the string. 
In fact, $t_e^{max}$ is an upper bound on the time of emission for rays which can be detected
at the moment $t_o$.

\begin{figure}[h]
\begin{center}
\includegraphics[height=8cm,width=10cm]{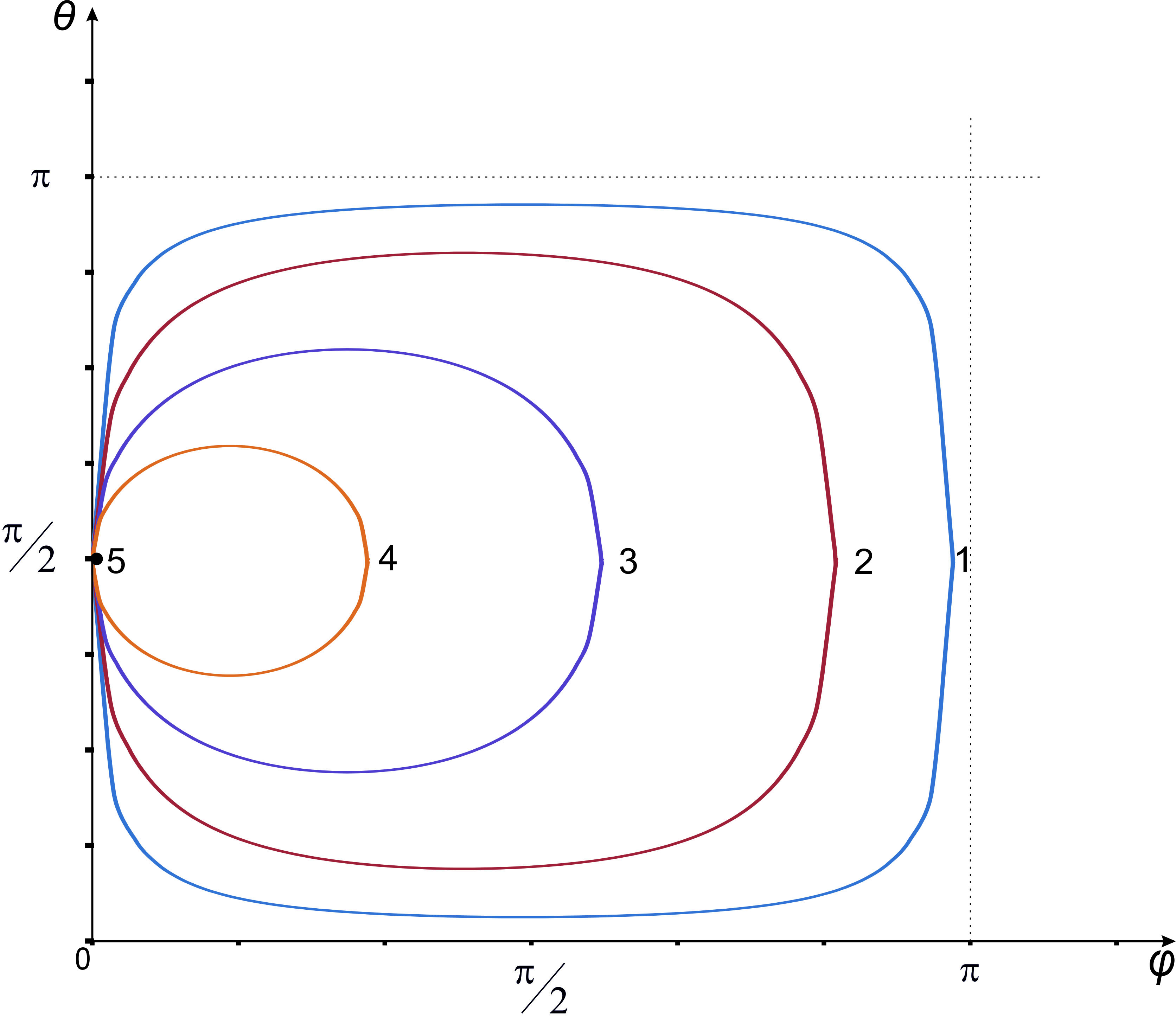}
\caption{\small{The evolution of the string image with time. An observer starts to see the image as a point at the moment it crosses the string horizon. Then the image 
looks as a closed curve which grows with time.}}
\label{F3}
\end{center}
\end{figure}

The string image grows with time, as is shown on Figure \ref{F3}. When the observer crosses the horizon
at $t_o=0$ ($\bar{\theta}=\pi/2$) the image is a point  with coordinates $\theta=\pi/2,\varphi=0$ (point 5 on Figure \ref{F3}). The curves on the Figure correspond to the following moments of observations: $t_o=0.5 a$ (4), $t_o=1.2 a$ (3),  $t_o=3 a$ (2), $t_o=10 a$ (1). 
If $t_o\gg a$ ($\sin\bar{\theta}=0$) the curve approaches a rectangular with vertexes at 4 points $(\theta=0,\varphi=0), (\theta=0,\varphi=\pi), (\theta=\pi,\varphi=0), (\theta=\pi,\varphi=\pi)$.

One can calculate the coordinate velocity of a typical point on a string image, say $(\theta=\pi/2, \varphi=\pi-2\bar{\theta})$, which allows one to estimate how fast the string image grow is. One has 
$$
{d\varphi \over dt_o}=-2{d\bar{\theta} \over dt_o}={2a \over a^2+t_o^2}~~.
$$

We can now consider an analogue of the Kaiser-Stebbins effect. Suppose that the string moves through a gas of photons. The radiation is isotropic and homogeneous in the observer's frame of reference (before crossing the horizon), frequencies of quanta are equal to $\omega$. From the point of view
of the 'right' observer, the MCS changes the frequencies of 'left' quanta
he can detect. These quanta
make 'left' part of the observer's light cone shown on Figure \ref{F1}. 

The observer sees 'left' rays as coming from the directions inside the string
image, Figure \ref{F2}. To see this let us take 'left' rays which come out from 
the string horizon with coordinate $y=-b$, where $b>0$.  The image of such rays
as seen by the observer is given by equation (\ref{5.3}), where $\bar{\theta}$
should be replaced with a parameter $\bar{\theta}'$ ($0<\bar{\theta}'<\pi/2$) 
determined from the relation
\begin{equation}\label{5.5}
\sin\bar{\theta}'={a' \over \sqrt{(a')^2+t_o^2}}~~,
\end{equation}
with $a'=a+b$. Since $\bar{\theta}'>\bar{\theta}$, the considered rays make a closed 
curve congruent to and lying inside the string image.

Thus, all photons with perturbed frequencies are seen as a spot on the sky whose boundary is the string image. The spot grows from a point and at late times it covers
the hemisphere $0<\theta <\pi$, $0<\varphi <\pi$.

Equation (\ref{2.8a}) allows one to find energies $\omega'$ of photons inside the spot. 
Let $(k')^\mu$ be a four-momentum of a 'left' quantum registered by a right observer,
$(k')^0=\omega'$, ${\bf k}'=\omega'{\bf n}$. With the help of (\ref{2.8a})
one gets
\begin{equation}\label{5.6}
\omega=f(\lambda, {\bf n}) \omega'~~,
\end{equation}
\begin{equation}\label{5.7}
f(\lambda, {\bf n})=\left(1+{\lambda^2 \over 2}\right)-{\lambda^2 \over 2}n^x-\lambda n^y~~.
\end{equation}
Function $f(\lambda, {\bf n})$ is positive by the construction (and this can be explicitly checked). From (\ref{5.6}) one finds the temperature anisotropy inside the spot
\begin{equation}\label{5.8}
{\delta T \over T}\equiv {\omega'- \omega \over \omega}
={1-f(\lambda, {\bf n}) \over f(\lambda, {\bf n})}\Theta({\bf n})~~,
\end{equation}
where $\Theta({\bf n})=1$ inside the spot and it is 0 outside.
At small energies of MCS, $EG \ll 1$, this formula simplifies to
\begin{equation}\label{5.9}
{\delta T \over T}=\lambda n_y\Theta({\bf n})~~.
\end{equation}
This is exactly the formula for temperature variations for a moving massive string.

In general, temperature variations (\ref{5.8}) caused by massless strings
look more complicated. For high energy MCS ($EG \simeq 1$, $\lambda \simeq 8\pi\simeq 25$) energies of photons in the spot are highly redshifted, $\omega'\simeq \omega \lambda^{-2}$.

A search for cosmic strings by studying CMB anisotropy is one of the promising methods for experimental detection of such objects \cite{Sazhina:2008xs}. If MCS existed in Nature the induced CMB anisotropy would be an important tool to discriminate them from massive strings.

\section{Summary and Discussion}\label{sum}

One of the aims of the present work was to reveal an operational meaning of MCS spacetime.  The MCS spacetime 
can be decomposed on two parts: below, $u<0$, and above,  $u>0$, the string horizon. Particles and rays move in these parts as in Minkowsky spacetime. The problem is how to describe their transitions across the horizon $u=0$.

The study of geodesics in Section \ref{reg} shows that one can introduce
two types of coordinate charts, $R$- and $L$-charts. $R$-charts are smooth everywhere except the cut, $u=0, y<0$, along of the horizon to the left from the string trajectory. 
Coordinates and velocities of geodesics which cross the cut
are transformed by the parabolic holonomy with respect to the $R$-chart. The holonomy
parameter $\lambda=8\pi GE$ is determined by the energy of the string.
On $R$-charts the 'right' trajectories behave smoothly across the horizon. Thus, the $R$-charts are convenient to study 
physical effects of MCS from the point of view of 'right' observers.
One can use here Cartesian coordinates with a cut on a half plane $u=0, y<0$.
The 'right' geodesics here are just straight lines.

The $L$-charts are dual to $R$-charts. They are smooth everywhere except the 
right cut on the horizon, $u=0, y>0$, where 'right' geodesics experience transformation
by the inverse holonomy. The $L$-charts are useful to study 
MCS effects by 'left' observers.

Note that the string horizon can be considered as a Cauchy hypersurface, where one sets initial data, coordinates and velocities, for geodesics lying at $u>0$. These
initial data are determined by incoming data. Depending on the chart initial and
incoming data differ by the holonomy either on the right or left parts of the horizon.

One can extend the above method to describe other physical phenomena.
The main interest is to classical and quantum fields $\phi$ on MCS spacetime (fibre bundles). Suppose that $\bar{\phi}(x)$ are incoming data  for a field on the 
string horizon. Then the initial data $\phi(x)$, say, in the $R$-chart look as follows:
\begin{equation}\label{6.1}
\phi(x)=\bar{\phi}(x)\mid_{u=0,y>0}~~,~~
\phi(x)=\bar{\phi}'(x')\mid_{u=0,y<0}~~,
\end{equation}
where $\bar{\phi}'$ and $x'$ are the field and its position
transformed on the horizon by the holonomy (compare with (\ref{3.4})). 
Equations (\ref{6.1}) also imply conditions on a number of field normal derivatives.
Concrete computations which can be done on the base of (\ref{6.1}) require 
field equations. We plan to return to this question in a future work. 

Some remarks are in order about the wake and Kaiser-Stebbins effects induced by MCS.
These effects show that MCS have distinguished features from the observational point of view in regime of high energies $EG\simeq 1$. 'Soft' MCS, $EG\ll 1$, behave as moving massive strings. 

Our results were obtained for MCS in a flat spacetime. Therefore, one should take care
in drawing conclusions how these effects look for MCS in the expanding Universe.
A study of MCS on curved manifolds would be an interesting research direction.

Finally, we did not mention any physical mechanisms which may result
in appearance of massless cosmic strings in the early Universe. Such mechanisms are not known, at least to the knowledge of the author.  

\bigskip
\bigskip
\bigskip

\noindent
{\bf Acknowledgement}

\bigskip

The author is grateful to  Andrei Zelnikov and Elena Timalina for a technical 
assistance with preparation of the paper.


\newpage

\begin{thebibliography}{}

\bibitem{Kibble:1976sj} T.W.B. Kibble, {\it Topology of Cosmic Domains and Strings},
J. Phys. {\bf A9} (1976) 1387-1398.

\bibitem{Vilenkin:2000jqa} A. Vilenkin, E.P. S. Shellard, {\it Cosmic Strings and Other Topological Defects}, Cambridge University Press, 2000. 

\bibitem{Stebbins:1987va}  A. Stebbins, {\it Cosmic Strings and the Microwave Sky. 1. Anisotropy from Moving Strings} 
Astrophys. J. {\bf 327} (1988) 584-614. 
 
\bibitem{Sazhina:2008xs} O.S. Sazhina, M.V. Sazhin, V.N. Sementsov, M. Capaccioli, G. Longo, G. Riccio, G. D'Angelo, {\it CMB Anisotropy Induced by a Moving Straight Cosmic String}, Conference: C08-05-23 Proceedings, 
e-Print: arXiv:0809.0992 [astro-ph].

\bibitem{Brandenberger:2013tr} R.H. Brandenberger, {\it Searching for Cosmic Strings in New Observational Windows}, Nucl. Phys. Proc. Suppl. {\bf 246-247} (2014) 45-57, 
e-Print: arXiv:1301.2856 [astro-ph.CO].

\bibitem{Barrabes:2002hn}   C. Barrabes, P.A. Hogan, W. Israel, {\it The Aichelburg-Sexl boost of domain walls and cosmic strings }, Phys.Rev. {\bf D66} (2002) 025032,
e-Print: gr-qc/0206021.
 
\bibitem{Schild:1976vq} A. Schild, {\it Classical Null Strings}, Phys. Rev. {\bf D16} (1977) 1722.

\bibitem{vandeMeent:2012gb}  M. van de Meent, {\it Geometry of massless cosmic strings}, Phys. Rev. {\bf D87} (2013) no.2, 025020, e-Print: arXiv:1211.4365 [gr-qc].


\bibitem{Fursaev:1995ef} D.V. Fursaev, S.N. Solodukhin, {\it On the description of the Riemannian geometry in the presence of conical defects}, Phys. Rev. {\bf D52} (1995) 2133-2143, e-Print: hep-th/9501127.

\end{thebibliography}
\end{document}